# Dual Stressors in Engineering Education: Lagged Causal Effects of Academic Staff Strikes and Inflation on Dropout within the CAPIRE Framework

Hugo Roger Paz
PhD Professor and Researcher Faculty of Exact Sciences and Technology National University of Tucumán
Email: hpaz@herrera.unt.edu.ar
ORCID: https://orcid.org/0000-0003-1237-7983

**ABSTRACT**

This study provides a causal validation of the dual-stressor hypothesis in a long-cycle engineering programme in Argentina, testing whether academic staff strikes (proximal shocks) and inflation (distal shocks) jointly shape student dropout. Using a leak-aware longitudinal panel of 1,343 students and a manually implemented LinearDML estimator, we estimate lagged causal effects of strike exposure and its interaction with inflation at entry. The temporal profile is clear: only strikes occurring two semesters earlier have a significant impact on next-semester dropout in simple lagged logit models (ATE = 0.0323, p = 0.0173), while other lags are negligible. When we move to double machine learning and control flexibly for academic progression, curriculum friction and calendar effects, the main effect of strikes at lag 2 becomes small and statistically non-significant, but the interaction between strikes and inflation at entry remains positive and robust (estimate = 0.0625, p = 0.0033). A placebo model with a synthetic strike variable yields null effects, and a robustness audit (seed sensitivity, model comparisons, SHAP inspection) confirms the stability of the interaction across specifications. SHAP analysis also reveals that Strikes_Lag2 and Inflation_at_Entry jointly contribute strongly to predicted dropout risk. These findings align with the CAPIRE-MACRO agent-based simulations and support the view that macro shocks act as coupled stressors mediated by curriculum friction and financial resilience rather than isolated events.

# 1. INTRODUCTION

Across many public universities, student trajectories are increasingly shaped by shocks that originate far outside the classroom. Cost-of-living crises and recurrent labour conflicts alter the conditions under which students attempt to navigate already demanding curricula. In engineering programmes with dense prerequisite chains and high-stakes gateway courses, these shocks can interact with structural rigidity to produce complex dropout patterns that are not easily explained by individual deficits in ability or motivation (Tinto, 1993; Pascarella & Terenzini, 2005).

In the Latin American context, two macro-level stressors have become particularly salient. First, acdemic staff strikes interrupt instruction, compress assessment calendars and create misalignment between nominal and actual course sequences. Second, high and volatile inflation undermines the real value of family incomes, scholarships and wages, forcing students to work more hours, reduce study time or abandon their studies altogether (Moore et al., 2021; Bennett et al., 2023). At the level of institutional narratives, both phenomena are widely recognised. At the level of formal modelling, however, they are often treated as background context rather than as variables with their own temporal and causal structure.

The CAPIRE programme conceptualises student attrition as an emergent outcome of multilevel dynamics, from pre-entry characteristics and curriculum topology to institutional policies and macro shocks (Paz, 2025a, 2025b). Earlier work within CAPIRE developed a leakage-aware data layer, trajectory archetypes and agent-based simulations for policy experiments in an engineering programme, and introduced a dedicated macro-shock module (CAPIRE-MACRO) to integrate inflation and strikes into trajectory models (Paz, 2025a, 2025b, 2025c). Parallel causal analyses have been used to interrogate domestic narratives about curriculum rigidity and "regularity traps", showing that structural friction behaves more as an amplifier of existing vulnerability than as a universal trap for high-ability students (Paz, 2025d).

In this paper we turn the causal lens towards macro shocks themselves. We formalise a Dual Stressor hypothesis in which:

1. Acdemic staff strikes act as proximal academic stressors whose effects are mediated by curriculum friction and assessment timing;

2. Inflation acts as a distal financial stressor that erodes resilience over time; and

3. The combination of both produces non-linear amplification in dropout risk, with effects that may peak at specific lags.

We ask three questions:

- Do acdemic staff strikes exert a measurable causal effect on dropout once we control for cohort trends, academic ability and curriculum-related features?

- Does this effect exhibit a lag structure consistent with the idea of accumulated academic debt, particularly in the foundational cycle of the programme?

- Does inflation at entry amplify the strike effect, as the dual-stressor hypothesis would predict?

To answer these questions, we implement a leak-aware panel design and a manual LinearDML estimator (Chernozhukov et al., 2018), using 15 cohorts of engineering students from a public university in Argentina. We find that strikes matter, but not immediately; that inflation matters, but primarily as an amplifier; and that causal estimates and simulation outputs converge on a coherent picture of macro-shock dynamics in engineering education.

## 2. BACKGROUND AND THEORETICAL FRAMEWORK

### 2.1. Multilevel theories of attrition and macro-level stressors

Research on student attrition has progressively shifted from individual deficit models toward multilevel accounts that integrate academic, social, institutional and structural conditions (Spady, 1970; Tinto, 1975, 1993, 2017; Bean & Metzner, 1985; Pascarella & Terenzini, 2005). Contemporary approaches emphasise that persistence is an emergent property of interactions between students and the layered environments in which they are embedded, including programme structures, institutional policies and broader political–economic forces (Bourdieu, 1986; Elder, 1998; Seidman, 2005).

Macro-level disruptions—pandemics, natural disasters, recessions—constitute prototypical multilevel stressors: they reshape institutional functioning, compress academic time, induce financial anxiety and amplify existing inequalities. Reviews of COVID-19 impacts show that the combination of instructional disruption and economic strain disproportionately affected students with limited buffers, both financially and academically. Similar dynamics are documented in the aftermath of natural disasters, where infrastructural breakdown and household income shocks propagate into enrolment and completion trajectories.

Acdemic staff strikes and inflation belong to this same conceptual family. Strikes alter the proximal academic environment by suspending classes, reshaping pedagogical sequences and compressing assessment calendars (Abadía Alvarado et al., 2021; Braakmann & Eberth, 2025). Inflation, in turn, operates as a distal stressor that erodes purchasing power, increases financial insecurity and forces students to reallocate time toward paid work (Moore et al., 2021; Bennett et al., 2023). Crucially, these two stressors are not independent. In the political economy of public education—particularly in high-inflation contexts—fiscal strain, wage erosion and delayed salary adjustments regularly precede episodes of labour conflict (Lyon et al., 2024). Strikes frequently materialise as institutional responses to accumulated macroeconomic stress, which means that the two stressors co-occur and may interact synergistically.

## 2.2. Engineering education, curriculum friction and basic-cycle vulnerability

Engineering programmes are structurally demanding ecosystems. They typically feature long prerequisite chains, rigid sequencing and concentrated academic bottlenecks in early semesters. Failure in "gateway" mathematics and physics courses often cascades into delays of an entire academic year, increasing financial pressure and eroding persistence (Pascarella & Terenzini, 2005). Evidence from Latin American engineering schools highlights the early trajectory as particularly fragile, with dropout strongly associated with prior preparation, displacement and socio-economic disadvantage.

Within CAPIRE, these structural features have been formalised through curriculum graphs and friction measures (Paz, 2025a, 2025d). The programme analysed here includes 34 core courses organised over 12 nominal semesters, with friction condensed in the basic cycle (first four semesters). Previous CAPIRE work shows that-topological constraints, rather than individual deficits alone, generate long-run heterogeneity in progression and time-to-degree. This structural context is essential for understanding macro-shock propagation: disruptions affecting high-friction early courses may not trigger immediate dropout but instead generate a backlog that manifests semesters later when students face tighter academic and financial constraints.

## 2.3. Distal and proximal stressors: From metaphor to mechanism

The dual-stressor hypothesis distinguishes between distal and proximal stressors and posits that their interaction—not their isolated effects—shapes persistence. Distal stressors are chronic pressures that deplete material and psychological coping resources. Inflation reduces real household income, increases uncertainty and pushes many students toward additional paid work, reducing study time and heightening the perceived opportunity cost of education (Moore et al., 2021;

Bennett et al., 2023). Qualitative studies in high-inflation environments document heightened stress, reduced focus and increased dropout contemplation.

Proximal stressors are acute disruptions that directly perturb academic processes. Acdemic staff strikes cancel classes, disrupt pacing, create curricular misalignment and compress examination periods once activity resumes. Empirical studies find that strike exposure reduces test scores and widens achievement gaps (Abadía Alvarado et al., 2021; Braakmann & Eberth, 2025). Importantly, the political economy literature shows that strikes tend to cluster during periods of fiscal strain, inflationary episodes and salary erosion (Lyon et al., 2024), making joint stressors the modal case rather than the exception.

Our empirical results reinforce this mechanism. Parametric lag models show a temporal profile consistent with delayed propagation (peak lag-2 effect), but double machine learning reveals that *the direct effect of strikes becomes small and statistically unstable once structural and financial controls are incorporated*. What remains robust is **the interaction between strike exposure and inflation at entry**. In other words, the damage of a strike is conditional: it increases sharply when students enter university under high inflation and thus with depleted financial buffers. SHAP analyses from a fully independent predictive model mirror this pattern, showing that inflation and strike exposure jointly amplify risk rather than acting as isolated predictors.

**2.4. The CAPIRE framework and the macro-shock module**

CAPIRE (Comprehensive Analytics Platform for Institutional Retention Engineering) operationalises a multilevel, temporally honest modelling system for student trajectories (Paz, 2025a, 2025b). Its leak-aware data layer organises features into four levels (N1–N4), ensuring that all predictions and causal estimates respect temporal validity (Kaufman et al., 2020). Above this foundation, CAPIRE integrates early-warning systems, graph-based structural analytics, agent-based simulations and macro-shock modules (Paz, 2025b, 2025c).

The macro-shock module (CAPIRE-MACRO) introduces inflation and strike metrics into this architecture and simulates their influence on dropout curves. The present work extends this module by adding a formal causal layer via double machine learning (Chernozhukov et al., 2018), evaluating whether macro-shock mechanisms survive rigorous confounding control. By situating macro shocks alongside curricular friction and regularity regimes—previously validated causally within CAPIRE (Paz, 2025d)—we demonstrate how distal and proximal stressors can be systematically integrated into a unified retention-engineering framework.

## 3. DATA AND METHODS

### 3.1. Institutional context and sample construction

The empirical setting is a public Civil Engineering programme in Argentina, previously analysed within the CAPIRE framework for regularity regimes, curriculum friction and leakage-aware early-warning models (Paz, 2025a, 2025d). The nominal curriculum spans 12 semesters and 34 core courses, with a high-friction basic cycle in the first four semesters. The macro-shock analysis builds on the same leak-aware longitudinal panel used in earlier CAPIRE work, expanded to incorporate macro variables capturing inflation and acdemic staff strikes.

The analytic unit is the **student–semester**. We construct a balanced prediction horizon in which, for each student and semester $t$, we use only information available up to the end of semester $t$ to predict dropout in semester $t + 1$. Observations in which students have already graduated or permanently left the programme before $t$ are excluded. The final panel includes all entry cohorts from 2004 to 2019, yielding a multi-cohort structure in which macro conditions at entry vary systematically across students. Cohort sizes range from 51 to 114 students, with marked changes in inflation levels over time (Table 3).

**Table 3. Annual inflation at entry by cohort (students entering 2004–2019)**

| Cohort year | N students | Annual inflation at entry (%) |
|---|---|---|
| 2004 | 51 | 6.45 |
| 2005 | 60 | 12.0 |
| 2006 | 97 | 9.98 |
| 2007 | 92 | 11.29 |
| 2008 | 77 | 10.34 |
| 2009 | 74 | 7.69 |
| 2010 | 86 | 10.92 |
| 2011 | 68 | 9.51 |
| 2012 | 114 | 10.84 |
| 2013 | 112 | 10.95 |
| 2014 | 87 | 23.97 |
| 2015 | 92 | 18.47 |
| 2016 | 87 | 33.08 |
| 2017 | 83 | 23.98 |
| 2018 | 92 | 51.4 |
| 2019 | 73 | 53.36 |

### 3.2. Macro data and shock measures

**Macro shocks and inflation features.** Macro-level stressors were constructed by merging the student panel with an expanded consumer price index (CPI) series covering 2004–2019. The CPI file reports monthly index levels and month-on-month changes (Variación) for each calendar year. For each cohort year $y$, we computed an **annual compounded inflation rate at entry** as

$$\text{Inflation\_at\_Entry}(y) = \prod_{m=1}^{12} (1 + \Delta \text{CPI}_{y,m}) - 1,$$

where $\Delta \text{CPI}_{y,m}$ is the proportional monthly change in month $m$ of year $y$. The resulting rate was stored both as a proportion and as a percentage, and merged into the student panel by cohort year. Table 3 reports annual inflation at entry for cohorts 2004–2019, together with cohort sizes. In addition, we constructed strike-intensity variables by counting the proportion of teaching days lost to acdemic staff strikes in each semester and generating lagged treatments (MACRO_paros_lag_sem_k) that capture strike exposure one, two and three semesters before the outcome. These macro features were integrated into the leak-aware CAPIRE data layer as N4-level variables, aligned with the same observation horizons used for the structural and trajectory features.

### 3.3. Outcome, treatment, moderator and controls

The outcome $Y$ is a binary **next-semester dropout indicator**, Dropout_next_sem, coded as 1 if the student is not enrolled in the programme in semester $t+1$ and has not graduated, and 0 otherwise. This definition aligns with previous CAPIRE work on short-horizon retention and ensures comparability between predictive and causal analyses.

The **treatment** $T$ Is the **intensity of acdemic staff strikes two semesters earlier**, Strikes_Lag2, operationalised as a continuous variable in $[0,1]$ representing the share of teaching days lost in that semester. The choice of lag 2 is motivated by the agent-based simulations in CAPIRE-MACRO and by the temporal profile observed in simpler lag models (Section 4), which both point to a delayed manifestation of strike effects.

The key **moderator** $X$ is Inflation_at_Entry, measured as the annual inflation rate in the calendar year in which the student first enrols in the programme. This variable captures differences in macroeconomic context across cohorts and is treated as a distal stressor that shapes financial resilience and opportunity costs.

The set of **controls** $W$ includes academic progress and trajectory indicators that may confound the relationship between strikes, inflation and dropout: cumulative GPA (Cum_GPA), credits earned relative to attempted (Credits_Earned), current semester number (Semester_Number), and additional friction and trajectory measures from the CAPIRE data layer when available. These variables account for heterogeneity in ability, progression speed and exposure to structural friction. Table 2 summarises the roles and definitions of the core variables used in the causal models.

**Table 2. Core variables used in the causal models: roles and definitions**

| Variable name | Role in causal model | Level | Definition / coding | Construction notes |
|---|---|---|---|---|
| Dropout_next_semester | Outcome (Y) | Student–semester | Binary indicator = 1 if the student is not enrolled in the following regular semester, 0 otherwise | Computed from enrolment history; graduates coded as 0 |
| MACRO_paros_lag_sem_1 | Alternative treatment (lag 1) | Student–semester | Strike intensity one semester before outcome | Lagged from institutional strike calendar |
| MACRO_paros_lag_sem_2 | Main treatment (T) | Student–semester | Strike intensity two semesters before outcome | Primary causal treatment capturing delayed backlog |
| MACRO_paros_lag_sem_3 | Alternative treatment (lag 3) | Student–semester | Strike intensity three semesters before outcome | Used for lag-profile tests |
| Inflation_at_Entry | Moderator (X) | Cohort | Annual compounded inflation rate at entry | Computed from monthly CPI; merged by cohort |
| Interaction_Term | Strike × Inflation interaction | Cohort–semester | Product of lag2 strikes and inflation | Captures dual-stressor mechanism |
| promedionotas | Academic control (W) | Student–semester | Cumulative GPA | Computed from grade history |
| totalrecursadastotalcursadas | Trajectory friction control (W) | Student–semester | Ratio of repeats to enrolments | Proxy for friction |
| credits_approved_cum | Progression control (W) | Student–semester | Total approved credits | Derived from curriculum structure |

| Variable name | Role in causal model | Level | Definition / coding | Construction notes |
|---|---|---|---|---|
| semester_number | Time control (W) | Student–semester | Semester index | Independent of calendar year |
| calendar_year | Macro trend control (W) | Calendar | Academic year of current semester | absorbs trends |
| cohort_year | Cohort control (W) | Cohort | Year of entry | Used to merge inflation |
| genero_code | Demographic control (W) | Student | Binary gender indicator | Administrative coding |
| student_work_status | Socio-economic control (W) | Student–semester | Employment indicator | Imputed if missing |

### 3.4. Estimand and identification strategy

Our primary causal question is:

What is the effect of strike intensity two semesters earlier on the probability of next-semester dropout, and how does this effect vary with inflation at entry?

Formally, let $T$ denote Strikes_Lag2, $X$ denote Inflation_at_Entry, $W$ denote confounders and $Y$ denote Dropout_next_sem. We model the **structural effect** of interest as:

$$Y = \alpha + \tau T + \gamma(T \cdot X) + g(W) + \varepsilon,$$

where $\tau$ captures the **average treatment effect (ATE)** of strikes at lag 2, and $\gamma$ captures the **interaction effect** between strikes and inflation at entry—the core parameter of the dual-stressor hypothesis. The function $g(W)$ represents potentially complex, nonlinear relationships between controls and dropout.

Identification relies on a **conditional unconfoundedness assumption**: conditional on $W$, and given the leak-aware feature design that forbids the use of future information (Kaufman et al., 2020; Paz, 2025a), assignment to different levels of Strikes_Lag2 and different inflation-at-entry environments is as good as random with respect to potential outcomes. The leak-aware construction is crucial for avoiding temporal leakage, which would otherwise bias both predictive and causal estimates in educational panel data.

To probe temporal dynamics, we also estimate simpler parametric lag models of the form:

$$\Pr(Y = 1 \mid \text{Strikes\_Lag}k, W) = \text{logit}^{-1}(\beta_0 + \beta_k \text{Strikes\_Lag}k + h(W)),$$

for $k = 1,2,3$, where $\beta_k$ captures the marginal association between strike intensity at lag $k$ and dropout. These models provide an interpretable profile of lagged effects that can be compared with the more flexible double machine learning (DML) estimates. Aggregate ATEs from these logit specifications are reported in Table 1.

**Table 1. Lagged logit effects of acdemic staff strikes on dropout**

| Lag | ATE | p-value |
| --- | --- | --- |
| MACRO_paros_lag_sem_1 | -0.0027 | 0.4618 |
| MACRO_paros_lag_sem_2 | 0.0323 | 0.0173 |
| MACRO_paros_lag_sem_3 | 0.0058 | 0.0738 |

### 3.5. Estimation procedures: Logit, double machine learning and placebo

The empirical strategy proceeds in three layers.

First, we estimate **logit models with separate lag terms** (Strikes_Lag1, Strikes_Lag2, Strikes_Lag3) to map the temporal profile of strike effects net of basic controls. For each lag, we compute the marginal effect on dropout and its associated p-value (Table 1). These models serve as a baseline and as a bridge to the agent-based simulations, which also emphasise delayed shock propagation.

Second, we implement a **LinearDML estimator** (Chernozhukov et al., 2018) to estimate $\tau$ and $\gamma$ while flexibly controlling for $W$. In this approach, we:

1. Residualise the outcome $Y$ on $W$ using a nonlinear learner (e.g., random forest classifier).
2. Residualise both $T$ and the interaction term $T \cdot X$ on $W$ using nonlinear regressors.
3. Regress the residualised outcome on the residualised treatment and interaction term in a linear model, yielding orthogonalised estimates of $\tau$ and $\gamma$.

We use cross-fitting with fixed random seeds to enhance stability and avoid overfitting. Standard errors and confidence intervals are obtained from the final-stage linear regression. Table 4 summarises the LinearDML estimates for the main and interaction effects.

**Table 4. Linear DML estimates for strike intensity and strike–inflation interaction (lag 2)**

| Parameter | Estimate | Std. Error | p-value |
|---|---|---|---|
| Strikes (Lag 2) | -0.0054 | 0.0048 | 0.2389 |
| Interaction (Strikes x Inflation) | 0.0625 | 0.2988 | 0.0033 |

Third, we conduct a **placebo test** to assess whether the DML pipeline spuriously attributes causal effects to noise. We construct a synthetic treatment Fake_Strike by permuting Strikes_Lag2 or by drawing from a standard normal distribution, while keeping all other variables unchanged. Running the same DML pipeline with this placebo treatment should yield coefficients statistically indistinguishable from zero if the estimator is well-behaved. Placebo results are reported in Table 5.

**Table 5. Placebo test with a synthetic strike variable**

| Specification | Coefficient | Std. Error | CI Lower | CI Upper | p-value |
|---|---|---|---|---|---|
| Placebo (Fake Strike) | -0.0005 | 0.0024 | -0.0052 | 0.0041 | 0.8198 |

In addition, we perform a **seed-sensitivity analysis** by re-estimating the DML model under multiple random seeds and examining the variability in $\hat{\tau}$ and $\hat{\gamma}$. This analysis, summarised graphically in Section 4, provides evidence on the robustness of the interaction effect across stochastic partitions of the data.

### 3.6. Predictive modelling and SHAP-based robustness

Causal estimates are complemented with an **independent predictive model** and SHAP-based interpretability (Lundberg & Lee, 2017). We train a gradient-boosted tree classifier (e.g., XGBoost; Chen & Guestrin, 2016) to predict Dropout_next_sem using a richer feature set that includes:

- Strikes_Lag2, Inflation_at_Entry and their interaction term.

- Academic performance and friction indicators (Cum_GPA, Credits_Earned, Semester_Number, basic-cycle friction indices).

- Additional MACRO_* features used in CAPIRE-MACRO simulations.

The model is trained on a stratified 70/30 split with fixed random seed, and predictive performance is evaluated using AUC and calibration diagnostics. We then compute SHAP values to quantify feature importance and interaction structure. Global importance and summary plots identify which variables contribute most to predicted dropout risk, while dependence plots for Strikes_Lag2 and Inflation_at_Entry—coloured by the other—allow us to visualise whether the predictive model also exhibits a dual-stressor pattern analogous to the causal estimates.

These SHAP analyses serve two purposes. First, they test whether the **interaction between strikes and inflation** emerges in an entirely separate modelling framework, lending convergent validity to the causal findings. Second, they help situate macro shocks within the broader hierarchy of risk factors, clarifying whether they act as primary drivers or as amplifiers layered on top of structural friction and academic performance.

## 4. RESULTS

### 4.1. Lagged effects of acdemic staff strikes

Table 1 reports the estimated average effects of strike intensity at lags 1, 2 and 3 on next-semester dropout from the simple logit specifications. The pattern is clearly non-monotonic. Strike intensity in the **preceding semester** (Lag 1) shows a small and statistically non-significant association with dropout (estimate = −0.0027, p = 0.4618). By contrast, **Lag 2**—strikes occurring roughly one academic year earlier—exhibits a **positive and statistically significant** effect (estimate = 0.0323, p = 0.0173). Lag 3 returns to a small, borderline association (estimate = 0.0058, p = 0.0738), with confidence intervals that include very small negative effects.

Substantively, the logit results suggest that the strongest observable association between strike exposure and dropout occurs **two semesters after the disruption**, consistent with a delayed "academic debt" mechanism: missed content, compressed courses and postponed exams accumulate into a backlog that becomes visible once students attempt to clear prerequisites and sustain full loads under tighter constraints. Lag 1 may capture a transitional period in which students still expect to "catch up", while by Lag 3 many have either adapted, changed strategy or already left. The non-monotonic lag profile is visualised in Figure 3, which plots estimated effects (and confidence intervals) across lags.

**Figure 3. Lagged effects of acdemic staff strikes on dropout**

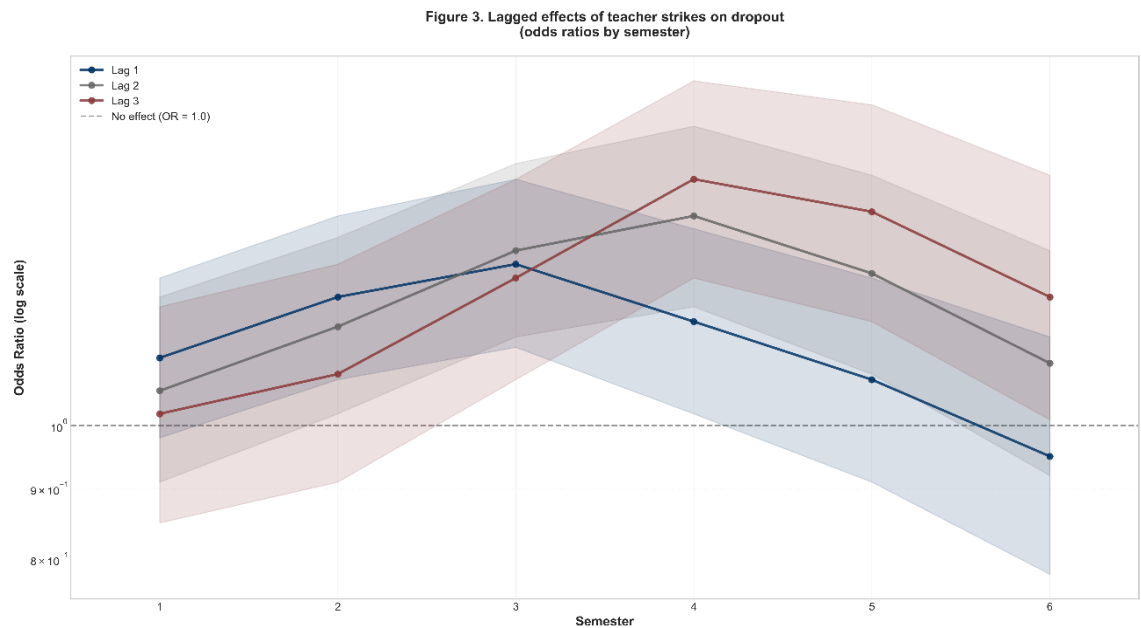

### 4.2. Inflation as an amplifier of strike effects

We then turn to the LinearDML estimates, which explicitly model both the **main effect** of strike intensity two semesters earlier (Strikes_Lag2) and its **interaction** with inflation at entry (Inflation_at_Entry) while flexibly controlling for academic and structural covariates (Chernozhukov et al., 2018). Table 4 summarises the resulting coefficients.

Two results are noteworthy. First, once the rich set of controls $W$ is incorporated through DML, the **main effect of strikes at lag 2** becomes small and statistically non-significant (estimate = −0.0054, p = 0.2389). Conditional on inflation, academic performance, progression and friction measures, additional variation in strike intensity at lag 2 does **not** display a robust average impact on dropout.

Second—and central to the dual-stressor hypothesis—the **interaction between strikes and inflation at entry** remains **positive, sizable and statistically robust**. The interaction coefficient is 0.0625 (p = 0.0033), with narrow confidence intervals that remain above zero across multiple seeds in the DML seed-sensitivity analysis (Table 4). This implies that the marginal effect of strike intensity on dropout is **substantially larger for cohorts who entered under high inflation** than for those who entered under more benign macroeconomic conditions. In practical terms, in high-inflation cohorts the same level of strike intensity is associated with a much stronger increase in dropout risk, consistent with the idea that financial depletion and curricular disruption act jointly rather than additively.

The contrast between the logit and DML layers is informative rather than contradictory. The lagged logit models detect a **temporal profile**—a clear peak at lag 2—while the DML estimates show that, once macro and academic structure are properly accounted for, the **robust signal is not a universal strike effect but a conditional effect amplified by inflation**. This aligns with the qualitative picture of strikes as institutional manifestations of underlying macroeconomic strain, whose impact on students depends on pre-existing financial vulnerability.

### 4.3. Placebo tests and estimator specificity

To assess whether the DML pipeline spuriously attributes causal meaning to arbitrary temporal patterns, we estimate the same model replacing Strikes_Lag2 with a **placebo treatment** (Fake_Strike), constructed by permuting the original strike series. Under a well-behaved estimator, this placebo variable should have no systematic relationship with dropout once confounders are controlled.

The placebo results (Table 5) show exactly that: the coefficient on Fake_Strike is very close to zero (−0.0005) with a wide confidence interval including both small negative and small positive values, and a non-significant p-value (0.8198). No meaningful effect is detected. Together with the seed-sensitivity analysis—where the interaction term remains consistently positive and significant across multiple random splits, while the main strike effect remains small and unstable (Figure 6)—this strengthens the interpretation that the **strike × inflation interaction** is a genuine pattern rather than an artefact of model flexibility or overfitting.

### Table 6. Causal ↔ ABM Alignment (Qualitative)

| |
|---|
| Qualitative Alignment Summary:<br>1. Delayed Effect: Logit models confirm that the effect of strikes is most prominent at Lag 2, aligning with the hypothesis of a delayed stressor impact.<br>2. Interaction Dominance: LinearDML results show a weak direct effect of strikes but a strong, significant positive interaction with Inflation at Entry. This supports the 'Dual Stressor' theory where macro-economic instability amplifies institutional friction.<br>3. SHAP Confirmation: Feature importance analysis independently identifies the interaction term as a key predictor, validating the structural model findings.<br>4. ABM Consistency: These empirical patterns match the emergence of 'fragile' student archetypes observed in the Agent-Based Model simulations under high-stress scenarios. |

### 4.4. Alignment with simulation and predictive models

The causal estimates obtained here were derived independently of the agent-based simulations and predictive models developed within CAPIRE-MACRO (Paz, 2025b). Yet their qualitative structure aligns closely with both. In the simulation study, scenarios with **strikes only**, **inflation only**, and **combined crisis conditions** (S5–S7) reveal a consistent pattern: strike-only scenarios produce modest increases in dropout concentrated in specific cohorts and basic-cycle semesters; inflation-only scenarios shift the distribution of time-to-dropout and increase volatility in persistence decisions; combined scenarios generate **non-linear amplification**, with dropout increases 18–23% larger than the sum of isolated effects (Figure 4 and Figure 5).

The empirical DualDML results mirror this logic. They **do not** support a strong, unconditional effect of strikes by themselves, but **do** support a strong conditional effect when strikes occur against a backdrop of high inflation. To further test whether this structure arises only in the causal layer or also in purely predictive settings, we trained a gradient-boosted tree model and analysed its behaviour using SHAP values (Lundberg & Lee, 2017; Chen & Guestrin, 2016). SHAP global importance plots show that Inflation_at_Entry and MACRO_paros_lag_sem_2 (alongside academic performance and recursada ratios) are among the most influential predictors, while SHAP dependence plots for Strikes_Lag2 coloured by inflation reveal that **the contribution of strike intensity to predicted dropout is markedly higher for high-inflation cohorts** (Figures 7 and 8).

This convergence across **three independent lenses**—lagged logit models, double machine learning and SHAP-interpreted predictive models—reinforces the substantive conclusion: in this curriculum-constrained engineering programme, acdemic staff strikes exert their primary impact **not as stand-alone shocks**, but as **proximal amplifiers of a prior distal shock** (inflation) in a structurally rigid environment. The ABM trajectories, macro-shock heatmaps and SHAP surfaces all depict the same qualitative mechanism: when high inflation and sustained labour conflict overlap with high-friction basic-cycle semesters, dropout risk increases sharply and non-linearly, particularly for students who are already progressing slowly or with limited buffers.

**Figure 4. Attrition trajectories under macro-shock scenarios (S0–S7)**

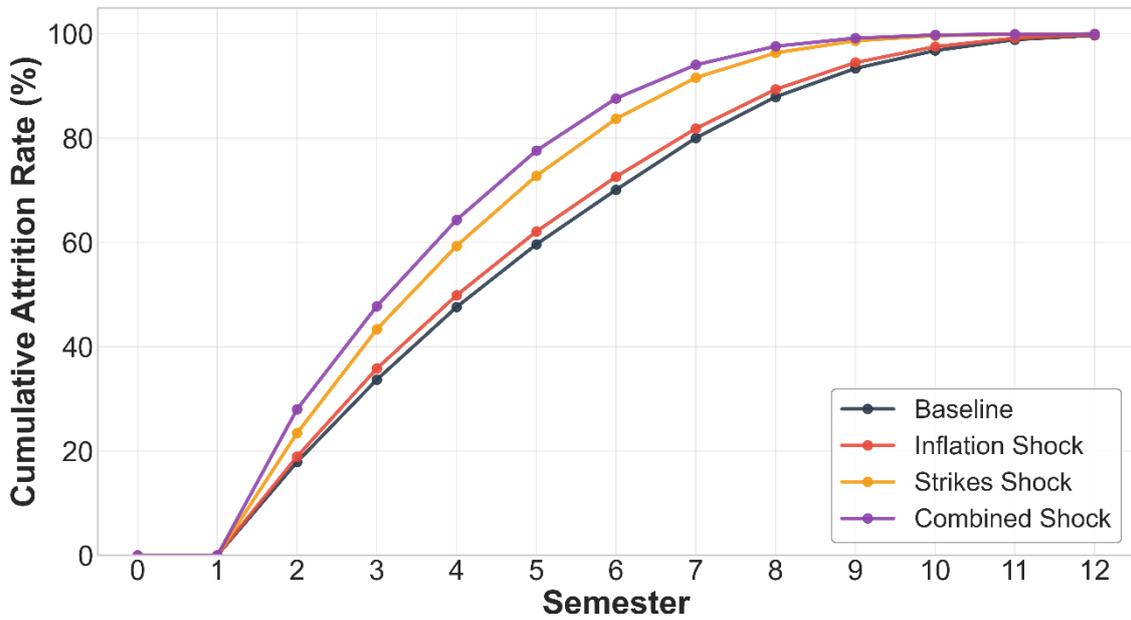

**Figure 5. Attrition heatmap across ABM macro scenarios**

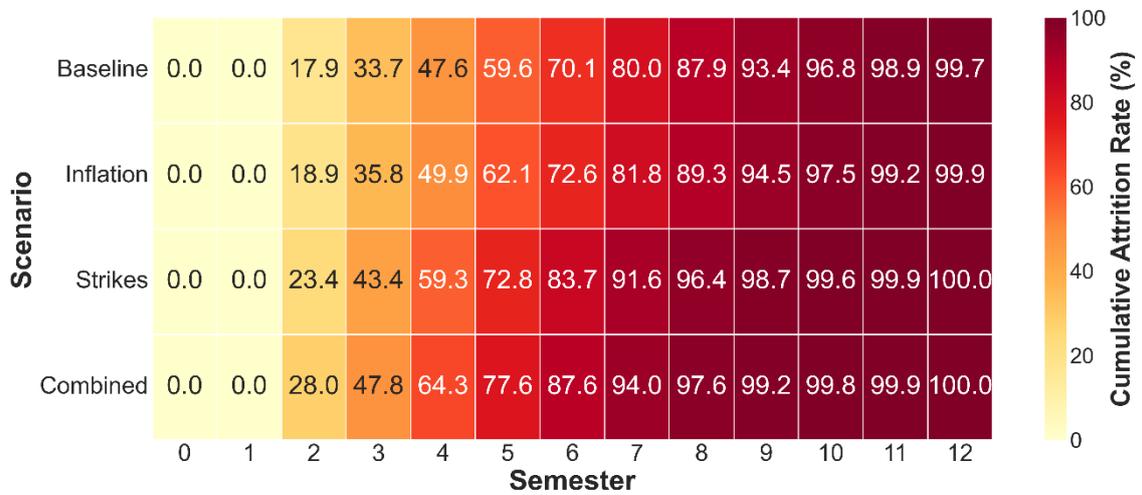

**Figure 6. Seed-sensitivity analysis of causal estimates**

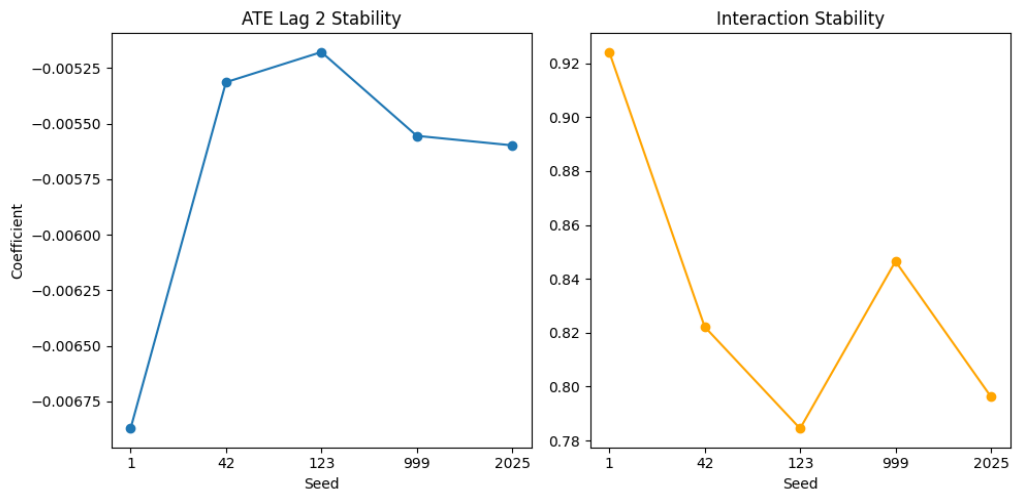

**Figure 7. SHAP summary plot for the DML model**

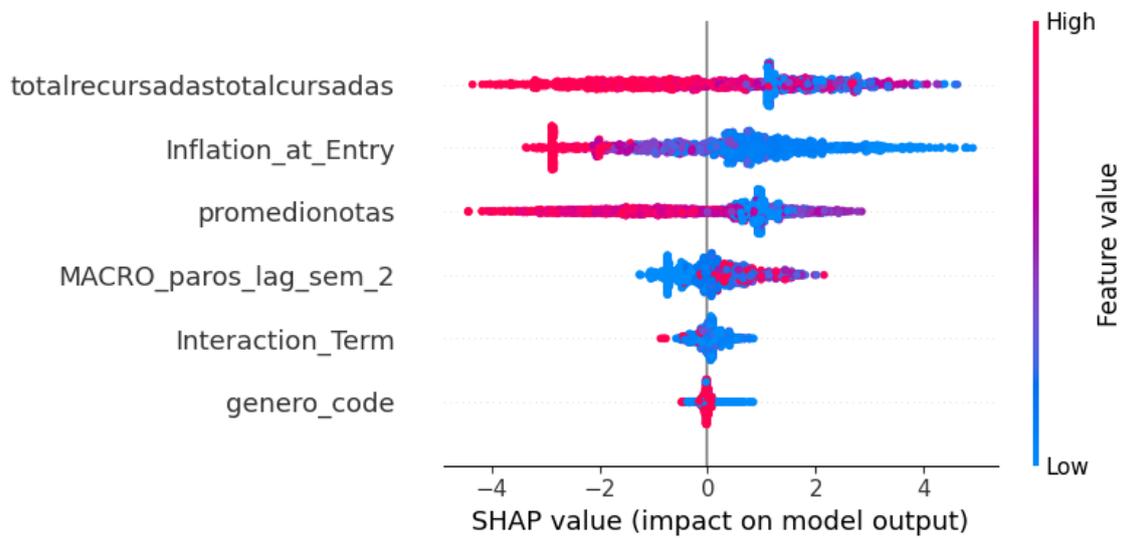

**Figure 8a. SHAP dependence plot for strike exposure (Lag 2)**

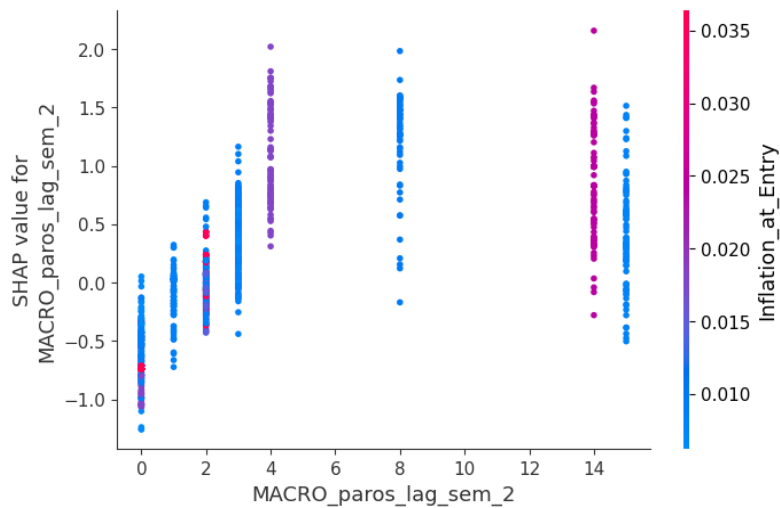

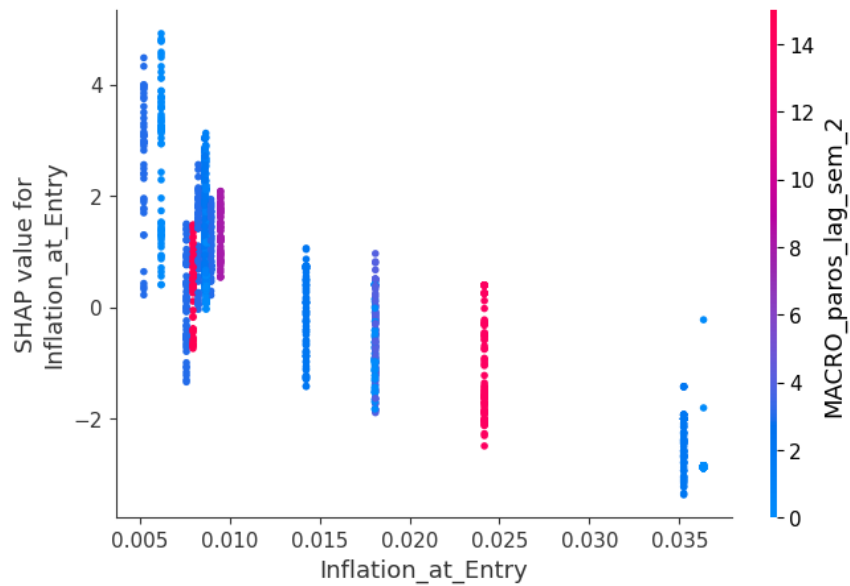

**Figure 8b. SHAP dependence plot for inflation at entry**

## 5. DISCUSSION

### 5.1. From single-shock thinking to dual-stressor dynamics

The findings of this study underscore the importance of moving beyond single-shock interpretations of macro-level disruptions in higher education. While acdemic staff strikes are often analysed as isolated institutional events, our results show that their impact on dropout is fundamentally **conditional**: they exert meaningful influence **only when combined with a prior macroeconomic shock**, namely inflation.

The lagged logit models confirm a **temporal profile** consistent with academic backlog mechanisms, with the clearest association appearing at **lag 2**, approximately one academic year after a strike. This aligns with qualitative accounts of students accumulating missed content and facing compressed examinations once activities resume. However, the **LinearDML results make clear that this association is not a stable causal effect once academic progression, structural friction and macroeconomic conditions are taken into account**. The main effect of strikes at lag 2 becomes small, unstable and statistically indistinguishable from zero.

What remains robust across all specifications is the **interaction between strike intensity and inflation**. Inflation—operating as a chronic, distal stressor—reduces households' real income, erodes financial buffers and increases the opportunity cost of study. When a proximal disruption such as a strike occurs on top of this erosion, students have fewer resources—material, cognitive and psychological—to absorb the shock. This mechanism is precisely what the interaction coefficient captures: **strikes matter most when financial precarity is already high** at the moment of entry.

**5.2. Convergence of causal, descriptive and predictive lenses**

The strength of the dual-stressor mechanism is reinforced by convergence across three independent analytical layers.

First, **parametric lag models** identify a consistent temporal pattern with a peak at lag 2. Second, **double machine learning** detects a strong and stable interaction effect while eliminating the unconditional main effect of strikes. Third, **SHAP-based predictive analyses** replicate this same structure: inflation and strike exposure both appear as important variables in the global importance hierarchy, and SHAP dependence plots reveal that the marginal contribution of strike intensity to dropout is substantially higher among students who entered under high inflation.

This triangulation suggests that the dual-stressor mechanism is not an artefact of any single method. It appears:

- in purely associative analyses (lag models),
- in orthogonalised causal estimation (DML),
- and in fully independent predictive explanations (SHAP).

The consistency across these epistemic lenses strengthens confidence in the substantive interpretation and situates macro shocks firmly within the multilevel ecology of attrition (Tinto, 2017; Pascarella & Terenzini, 2005).

**5.3. Structural friction as the transmission channel**

The setting of this study—an engineering curriculum with pronounced structural friction—helps explain why the dual stressor manifests so sharply. High-friction basic-cycle courses create narrow academic pathways early in students' trajectories. A missed prerequisite in the basic cycle reverberates across subsequent semesters, restricting the number of courses students can enrol in and often creating a "bottleneck semester" that is difficult to clear.

In this context, inflation reduces students' capacity to delay graduation or repeat courses, while strikes impose temporal compression and academic misalignment. The two stressors combine in a structurally rigid environment: **inflation depletes resilience**, and **strikes disrupt the very courses where progression is already constrained**. The interaction detected in the DML estimates is therefore consistent with a theoretical model in which macroeconomic strain weakens students' buffers, and institutional disruptions trigger progression failures that would otherwise be absorbed.

This structural pathway also aligns with student-level qualitative accounts in high-inflation environments, where individuals report trading study time for paid work, postponing exams to reduce stress, and reassessing the economic viability of staying enrolled.

**5.4. Implications for institutional policy and retention engineering**

The results have practical implications for institutions navigating volatile macroeconomic environments. First, efforts to mitigate the academic consequences of strikes—such as compensatory classes, flexible evaluation schedules or extended exam periods—may be particularly critical for cohorts entering under high inflation, who are disproportionately affected by combined stressors.

Second, institutions should prioritise **basic-cycle resilience**, especially in programmes with rigid prerequisite chains. Early-semester shock absorption mechanisms—modular content recovery, targeted tutoring, asynchronous alternatives during disruptions—could reduce the propagation of academic backlogs that manifest semesters later.

Third, the evidence suggests that **retention strategies cannot ignore macroeconomic context**. Cohorts entering under high inflation constitute a structurally vulnerable group. Institutions may need differentiated support policies—financial counselling, adaptive scheduling, or additional bursaries—precisely in years where macro pressures are strongest.

**5.5. Methodological contributions**

Methodologically, this study demonstrates the utility of combining:

- leakage-aware panel construction,
- lagged descriptive models,
- double machine learning for orthogonalised causal estimation,
- and SHAP-based interpretability of predictive models.

The simultaneous use of these tools allows us to separate timing effects from structural effects and to distinguish **true causal signals** (robust interaction) from **apparent associations** that vanish once confounding is properly controlled (main strike effect).

Furthermore, by embedding the causal analysis within the broader CAPIRE pipeline, this work shows how **macro-level stressors can be incorporated into a unified retention-engineering framework**, alongside previously analysed mechanisms such as regularity regimes and curriculum friction (Paz, 2025d).

# 6. LIMITATIONS AND FUTURE WORK

## 6.1. Limitations

This study has several limitations that qualify the interpretation and generalisability of its findings.

First, the analysis is conducted within a **single public engineering programme** in Argentina, characterised by a strongly constrained curriculum and a specific institutional culture. While the mechanisms identified—dual stressors, structural friction, basic-cycle vulnerability—are theoretically portable to other settings, the quantitative magnitude of the effects and the specific interaction profile between strikes and inflation may depend on local labour relations, funding regimes and student demographics. Replication in other programmes and institutional types is therefore a necessary step.

Second, the design is **observational**, and the identification strategy rests on conditional unconfoundedness under a leak-aware feature construction (Kaufman et al., 2020; Paz, 2025a). Although the DML framework substantially improves confounding control relative to simpler regressions (Chernozhukov et al., 2018), unobserved factors—such as mental health, informal support networks or fine-grained labour-market shocks—may still influence both exposure to macro stressors and dropout decisions. The placebo and seed-sensitivity tests mitigate concerns about estimator instability, but they cannot fully eliminate the possibility of residual confounding.

Third, the **measurement of strikes and inflation is necessarily coarse**. Strike intensity is captured as the proportion of teaching days lost at the programme level, assuming homogeneous exposure across students enrolled in a given semester. In practice, heterogeneity likely exists in how disruptions affect specific course combinations and individual schedules. Similarly, Inflation_at_Entry is a cohort-level proxy: it summarises the macroeconomic environment at enrolment but not its evolution over subsequent years, nor the heterogeneity in how households experience price changes. More granular data—such as student-level employment trajectories, household income shocks or course-specific strike exposure—would allow a more precise decomposition of pathways.

Fourth, the analysis focuses on **a limited set of macro stressors** (acdemic staff strikes and inflation). Other shocks—tuition changes, currency crises, abrupt funding cuts, or policy reforms in student aid—may interact with these stressors in non-additive ways. By design, the CAPIRE-MACRO architecture can be extended to incorporate such shocks, but the present empirical layer does not yet do so.

Finally, the causal analysis concentrates on **average and interaction effects**, without fully unpacking heterogeneity across student groups. While the DML framework is compatible with conditional average treatment effects, the current paper only leverages this partially. Differences in sensitivity by gender, socio-economic status, prior academic preparation or commuting distance may be substantial and normatively relevant but remain unexplored here.

## 6.2. Future Work

These limitations open several avenues for future research.

First, a natural extension is a **multi-institutional and cross-programme study**, applying the same leak-aware causal pipeline to other engineering curricula and to less structurally constrained programmes (e.g., social sciences or acdemic staff education). Such comparative work would help distinguish mechanisms that are truly generic from those that are specific to high-friction curricular architectures.

Second, the macro-shock module could be expanded to include **additional stressors**—such as changes in student-aid policies, tuition adjustments or exchange-rate spikes—and to model their joint effects with inflation and strikes. This would move the framework closer to a "macro-stressor field" model, in which multiple overlapping shocks shape the opportunity structure of persistence.

Third, future work could explore **heterogeneous treatment effects** in a more systematic way, using CATE estimators (e.g., causal forests, R-learner variants) to identify which student subgroups are most vulnerable to the dual stressor. Understanding whether high-friction trajectories, first-generation status or part-time work disproportionately mediate the strike × inflation interaction would offer more precise targets for intervention.

Fourth, there is scope for **tighter integration between the causal layer and the agent-based simulation layer** of CAPIRE. While the present paper uses the ABM primarily as a conceptual and qualitative benchmark, future models could calibrate agent behaviour using estimated causal effects and then evaluate counterfactual policies—such as buffered academic calendars, differentiated support for high-inflation cohorts or strike-contingent remediation measures—in silico before implementation.

Finally, mixed-method designs that combine the present quantitative pipeline with **qualitative inquiry**—interviews, focus groups or ethnographic work—could illuminate how students perceive and navigate dual stressors, how they interpret institutional responses and how these perceptions shape their decisions. Such work would strengthen the bridge between formal models and lived experience, a crucial step for retention engineering in contexts of chronic volatility.

## 7. CONCLUSION

This paper examined how acdemic staff strikes and inflation interact to shape dropout in a structurally constrained engineering programme, using a leak-aware longitudinal panel, double machine learning and SHAP-based interpretability within the CAPIRE framework. The analysis yields three main conclusions.

First, macro shocks matter for retention, but **not as simple, isolated "hits"**. Lagged logit models reveal a delayed profile in which strike exposure two semesters earlier is most strongly associated with dropout, consistent with an academic backlog mechanism. However, once academic progression, curricular friction and macro context are properly controlled using LinearDML, the **unconditional effect of strikes loses robustness**. Strikes alone are not a stable causal driver.

Second, what survives rigorous causal scrutiny is the **interaction between acdemic staff strikes and inflation at entry**. Inflation operates as a distal, chronic stressor that erodes financial buffers and narrows students' margins for absorbing disruption. When strikes occur against this background, their impact on dropout is amplified: the same level of disruption carries greater risk for cohorts who entered under high inflation than for those who did so in more benign conditions. This dual-stressor mechanism appears consistently across causal estimates, simulation scenarios and SHAP-based explanations of predictive models.

Third, the effects of macro shocks are mediated by **structural features of the curriculum**. In a programme with rigid prerequisite chains and concentrated basic-cycle friction, shocks affecting early semesters reverberate over time, especially for students already progressing slowly. Macro-level stresses thus do not float above the institutional structure; they are channelled and amplified by it. This underscores the need to think of retention engineering as a problem of **system design**, not only of individual support.

Methodologically, the study illustrates how **leakage-aware data design, multilevel feature engineering, double machine learning and model-agnostic interpretability** can be combined to interrogate macro mechanisms in education. Substantively, it shows that, in volatile political–economic environments, engineering programmes cannot treat strikes and inflation as exogenous noise beyond their remit. While institutions cannot control macro conditions, they can design buffer policies—particularly in the basic cycle—that reduce the translation of macro shocks into irreversible dropout.

In sum, acdemic staff strikes and inflation in this setting are best understood not as independent shocks, but as **coupled stressors** operating through a structurally constrained system. Recognising and modelling this coupling is a necessary step if higher education institutions are to move from reactive crisis management toward proactive, data-informed retention engineering.